\documentclass[preprint]{aastex}
\usepackage{graphicx}
\usepackage{natbib}
\newcommand{\HeII}{He~{\sc ii}}
\newcommand{\SiVII}{Si~{\sc vii}}
\newcommand{\MgVI}{Mg~{\sc vi}}
\newcommand{\MgVII}{Mg~{\sc vii}}
\newcommand{\NeVIII}{Ne~{\sc viii}}
\newcommand{\SiX}{Si~{\sc x}}

\newcommand{\FeX}{Fe~{\sc x}}
\newcommand{\FeXI}{Fe~{\sc xi}}

\newcommand{\FeXII}{Fe~{\sc xii}}
\newcommand{\FeXIII}{Fe~{\sc xiii}}
\newcommand{\FeXIV}{Fe~{\sc xiv}}
\newcommand{\FeXV}{Fe~{\sc xv}}
\newcommand{\FeXVI}{Fe~{\sc xvi}}
\newcommand{\CaXVII}{Ca~{\sc xvii}}
\newcommand{\kms}{km~s$^{-1}$}

\newcommand{\hinode}{{\it Hinode}}

\begin{document}

  \title{EIS/Hinode observations of Doppler flow seen through the 40\arcsec\ wide slit}

\author{D.E. Innes\altaffilmark{1}, R. Attie\altaffilmark{1}, H. Hara\altaffilmark{2} and M.S. Madjarska\altaffilmark{1}}

   \altaffiltext{1} {Max-Planck Institut f\"{u}r Sonnensystemforschung,
  37191 Katlenburg-Lindau, Germany}
  \email{innes@mps.mpg.de}
  \altaffiltext{2}{National Astronomical Observatory, Mitaka, Tokyo 181-8588, Japan }

\begin{abstract}
The Extreme ultraviolet Imaging Spectrometer (EIS) on board \hinode\ is the
first solar telescope to obtain wide slit spectral images
that can be used for detecting Doppler flows in transition region and coronal lines
 on the Sun and to relate them to their surrounding small scale dynamics.
We select EIS lines covering the
temperature range $6\times10^4$~K to $2\times10^6$~K that give spectrally pure
images of the Sun with the 40\arcsec\ slit. In these images Doppler
shifts are seen as horizontal brightenings. Inside the image it is difficult to
distinguish shifts from horizontal structures but emission beyond the
image edge can be unambiguously identified as a line shift in several
 lines separated
from others on their blue or red side by more than the width of the
spectrometer slit (40 pixels). In the blue wing of \HeII, we find a large
number of events with properties (size and lifetime) similar to the
well-studied explosive events seen in the ultraviolet spectral range.
 Comparison with X-Ray Telescope (XRT) images shows many Doppler
 shift events at the footpoints of small X-ray loops.
The most spectacular event observed showed a strong blue shift in
 transition region and lower corona lines from a small X-ray spot that lasted
 less than 7~min.
 The emission appears to be near
  a cool coronal
loop connecting an X-ray bright point
to an adjacent region of quiet Sun. The width of the emission
 implies a line-of-sight
velocity of 220~\kms. In addition, we show an example of an \FeXV\ shift with
a velocity about 120~\kms, coming from what looks like a narrow loop leg connecting
a small X-ray brightening to a larger region of
X-ray emission.

\end{abstract}



\section{Introduction}
\begin{figure}
\centering
\includegraphics[width=12 cm]{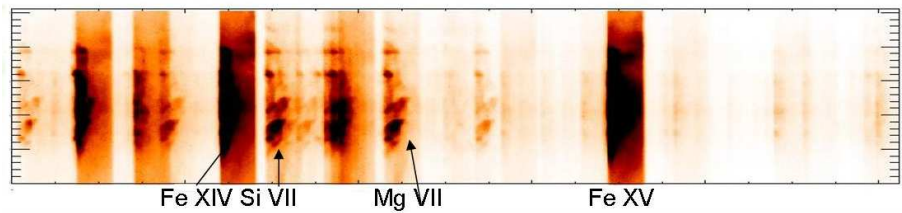}
\caption{A 40\arcsec\ wide slit spectrum of a small active region,
showing well separated and blended lines. Several of the lines discussed in this
paper are marked.}
    \label{slots}
   \end{figure}

The Extreme ultraviolet Imaging Spectrometer \citep[EIS;][]{Culhane07} on
\hinode\ obtains images and spectra of many transition region and
coronal lines in the wavelength ranges $170-211$~\AA\ and $246-292$~\AA.
The EIS
wide slits are an interesting compromise between narrow slit spectra
where the time to raster across
a particular structure is often longer than the lifetime of the structure
and filter images which give no direct measurement of flow velocities.
The 40\arcsec\ wide slit
 provides overlapping spectra from the observed  40\arcsec\
 wide region of the Sun. Images in different lines overlap if the lines are separated
by less than the 40 pixel width of the slit image
times the wavelength plate scale, 0.0223 \AA/pixel.
Figure~\ref{slots}, shows a 40\arcsec\ wide slit spectrum of a small active region
for the wavelength region $268-292$~\AA. Isolated lines produce well-defined
40\arcsec\ wide images.

Within the EIS spectrum there are several relatively
strong isolated lines that can be used to study the dynamics of the
outer atmosphere.
\citet{Hansteen07} report on rapid temporal
variations in quiet Sun \HeII\ 256~\AA\ and \FeXII\ 195~\AA\ wide slit
emission features observed with a cadence of 30~s.
They also mention the detection of many \HeII\ blue shifts
in quiet Sun narrow slit rasters observed after the wide slit sequence.

In this paper we show how
the detection of \HeII\ blue shifts and large-scale
temporal variations, can be made simultaneously with wide slit
observations.
The key is that the emission beyond the edge
of the 40\arcsec\ main line image
is either line broadening/Doppler shifts from
the main line or emission  in a neighbouring line.
Figure~\ref{slots} demonstrates both effects.
\FeXV\ and \FeXIV\ (labelled) are both well separated lines and the images are basically
straight along both edges. On the edge that cuts
the active region, the emission from both lines bulges slightly
because the bright active region lines are broader.
A closer inspection of the emission shows that both lines have the same
structure along the edge.
On the other hand, the strong unmarked line on the
left is mainly \FeXIV\ 270.5~\AA, but there is clearly a blend producing
extensions on the left hand side.
 Positions of the projections coincide with bright \MgVII\ and
\SiVII\ because the blend is the lower
coronal line \MgVI\ 270.4~\AA.

 When
the Sun's structure is known from images in isolated lines,
 it is possible to work out from the spectrum whether neighbouring lines are
 expected to produce emission.
If there is no spectral line that can cause the emission, it is
 very likely due to Doppler shifts/broadening in the image line.
Thus the EIS 40\arcsec\ wide slit
can be used to obtain both images
of the dynamics over a wide range of temperatures, and
Doppler shifts from structures on the edge of the images.

Inside the image,
Doppler broadened lines appear as horizontal streaks which are difficult to
distinguish from a jet-like structure.
Comparison with filter images is very important in order to
positively identify shifts inside the image. Simultaneous images obtained by
other instruments on \hinode\ can, in principle, be used to help identify
shifts but there will always be some uncertainty because the filters to not
detect exactly the same plasma.

In this letter we show examples of Doppler shift events seen in the quiet Sun
with the EIS 40\arcsec\ wide slit. Comparisons are made with X-ray images
obtained by the X-Ray Telescope \citep[XRT;][]{Golub07} in order to show the
relationship to hot loop structures. Further work will report on detailed
analyses with data from the Solar Optical Telescope \citep[SOT;][]{Tsuneta08}
on \hinode\ and the Solar Ultraviolet Measurements of Emitted Radiation
\citep[SUMER;][]{Wetal95} spectrograph on SoHO.

\section{Observations}
Observations are shown of regions of quiet Sun near disk center on 10 April
2007 and 7 November 2007, and near a small active region on the 15 November
2007. During the April observations, EIS observed with the 40\arcsec\ slit
and a cadence of 30~s at a single position. During the November runs EIS
alternated between two slit positions separated by 40\arcsec\ with an exposure time
of 30~s and cadence
of 90~s. In April, windows centered on the lines given in
Table~\ref{tab_lines} were recorded.
To
keep within the telemetry rate, JPEG lossy compression (Q=98) was used. In
November only the ones marked with an asterisk in Table~\ref{tab_lines}
 were recorded and lossless compression was
used.
For all observations the window height was 512\arcsec\ and for most lines
the window width was 56 wavelength pixels. The windows around the EIS core
lines \HeII, \FeXII\ and \FeXV\ were 64 pixels wide to
ensure at least 8 pixels either side of the image.
\begin{table}
\caption{Spectral lines in the center of each window during the 10 April 2007
observations. The ones
in bold are illustrated in Figure~\ref{slotprofs}. Those with an asterisk
are separated from other strong lines and produce spectrally pure images. They
were observed in November 2007.}
\label{tab_lines}
\begin{tabular}{l c c c c}
Ion & \multicolumn{3}{c}{Wavelengths} & Log $T$ (K)\\
\hline
\HeII & {\bf 256.32}$^*$ &  & &4.7\\
\MgVII &{\bf 278.39}$^*$ & 280.75 & & 5.8 \\
\SiVII & {\bf 275.35}$^*$ & & & 5.8\\
\FeX &184.54 & 190.04 & & 6.0\\
\SiX & 258.37$^*$ & {\bf 261.08}$^*$  & {\bf 272.01}$^*$ & 6.1\\
\FeXI &180.40$^*$ &188.25 & & 6.1 \\
\FeXII &{\bf 195.12}$^*$ & & & 6.1\\
\FeXIII &202.08 &{\bf 203.86$^*$} & & 6.2\\
\FeXIV & 264.88 & {\bf 274.20}$^*$ & & 6.3 \\
\FeXV &{\bf 284.16}$^*$& & & 6.3\\
\FeXVI &263.02$^*$ & & & 6.4\\
\CaXVII &192.82 & & & 6.7\\
\end{tabular}
\end{table}

\begin{figure}
\centering
\includegraphics[width=8.5 cm]{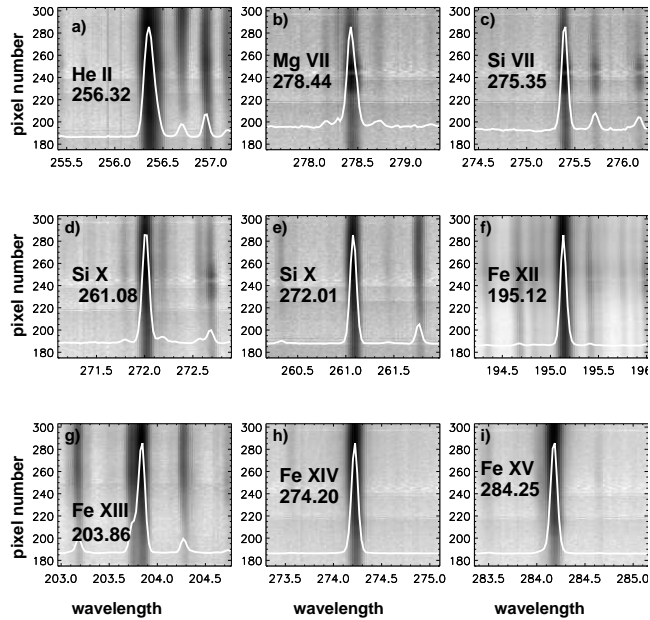}
\caption{The 80 pixel wide spectral region around selected strong isolated lines
giving the purest 40\arcsec\ slit images. The background of each frame is a stigmatic
spectral image taken with the 2\arcsec\ slit across a small active region on 3 April 2007.
The images are negatives and bright lines appear dark. Higher temperature
lines are stronger at the top of the images. Superimposed are
the average line profiles along the length of the slit.}
    \label{slotprofs}
   \end{figure}

The 80 pixel spectral regions around nine of the strongest isolated
 lines
are shown in Figure~\ref{slotprofs}.
The cleanest line is \FeXIV\ 274.20~\AA.
The two lines, \HeII\ 256.32~\AA\ and \SiVII\ 275.35~\AA\ are clean on their blue side.
Other lines, \FeXV\ 284.25~\AA\ and  \FeXIII\ 203.86~\AA\
are clean for more than $15\times0.0223$~\AA\ or 400 \kms\ to one or both sides.
These lines are  suitable for obtaining Doppler shifts at the edge
of the 40\arcsec\ images.
The \SiX\ 272.01~\AA\ appears unblended in Figure~\ref{slotprofs}, but
it is known to be blended with several transition region
lines \citep{Young07a},
so is only useful as a check on what one sees in other lines.

In order to assess this method of measuring
velocities, we constructed pseudo wide slit images from rasters of 2\arcsec\
slit observations. Line profiles with significant extensions two pixels
 from the line center, could be identified as Doppler shifts at the window
edge. This corresponds to 52 \kms\ at \HeII~256~\AA\ and 48~\kms\ at \FeXV~284~\AA.

The EIS images suffer from misalignment due to spacecraft jitter and internal
drift due to EIS temperature changes. The former changes the field-of-view,
and the latter the shift of the image on the CCD.
  The instrument
movement due to spacecraft jitter in the sun-x (spectral)
and sun-y directions was corrected by first coaligning all
images in a particular spectral window to the average for that window. This
was done to assess the best window for obtaining the instrumental
shifts. The \HeII, \SiVII\ and \MgVII\ coalignment parameters were similar and
appeared to successfully remove the instrument movement. Therefore these were
applied to all images in the same sequence. The maximum shift is 1.5 pixels in the
sun-y direction and less than 1 pixel in the sun-x direction. To investigate
emission just beyond the window edge, the movement of the window edge rather
than the solar features were taken into account.
Over a typical one hour
observing period the window edge drifted less than 1 pixel on the detector.
It is important to note therefore that images used for computing the
Doppler shifts beyond the edge have a slightly different co-alignment than those
that show the structural changes.

In this paper, the EIS wide slit images are presented such that the sun-y co-ordinate
increases towards the top and the sun-x co-ordinate
increases to the right of the images. The wavelength scale then
decreases to the right, so that blue shifts are seen beyond the right edge of the
images.

Simultaneous XRT images were obtained in April with the Al-mesh filter and a
cadence of 20~s and in November with the C-poly filter and a cadence of 30~s.
The data were corrected using the standard solarsoft XRT software.
Co-alignment of the EIS and XRT images was achieved by cross-correlating
common \FeXV\ and X-ray features.

\section{Dynamics}

\subsection{\HeII\ Explosive Events}

Explosive events are characterized by Doppler broadened line wings in transition
region lines. They have been well studied with HRTS \citep{BB83, DBB89}
and SUMER \citep{IIAW97b, IBGW97a, Cetal98, NIS04} ultraviolet spectra.
In the quiet Sun, most strong events are seen in lines
formed in the transition
region  with formation temperature $6\times10^4-3\times10^5$~K \citep{WEMW02}.
In active regions strong explosive events have also been reported in lower
corona lines such as \NeVIII\ with formation temperature around
$7\times10^5$~K \citep{Wilhelm98}.

There are several transition region and lower corona lines in the EIS
 wavelength range \citep{Young07b}. The three lines
 \HeII\ 256.32~\AA, \MgVII\ 278.39~\AA, and \SiVII\ 275.35~\AA\ all give sharp wide slit
 images. \HeII\ is ideal for detecting explosive events at the image edge
 because it is formed around $6\times10^4$~K and there are no lines on its
 blue side that will overlap with the Doppler shifted emission.
 The red side of  the \HeII\ line is
 blended and overlaps with images produced by the
 strong coronal lines \SiX\ 256.37~\AA, \FeXIII\ 256.42~\AA, \FeXII\ 256.41~\AA,
 and \FeXII\ 256.94~\AA, so \HeII\ blue shifts are more easily detectable.
 The best line for shifts in the lower corona is \SiVII\ 275.35~\AA. It is
 clean on the blue side and  for 14 pixels to the red where there is
   a second weaker \SiVII.
 The \MgVII\ 278.39~\AA\ shows blending with coronal lines
 in Figure~\ref{slotprofs} and is known to be blended with \SiVII\
 \citep{Young07b}, so may not always be easy to interpret in wide slit images.

\begin{figure}
\centering
\includegraphics[width=9.0 cm]{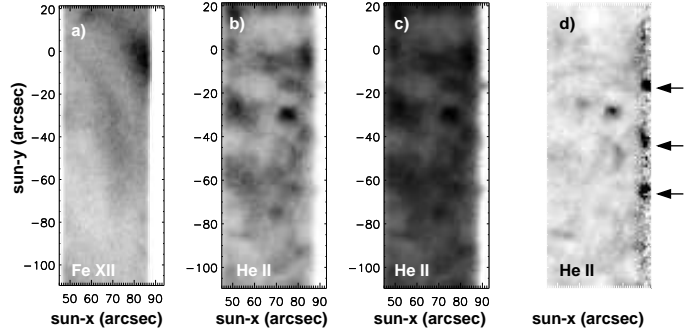}
\caption{Explosive events seen in \HeII\ in the quiet Sun images taken at
21:45:57~UT on 10 April 2007, in \FeXII\ 195~\AA\ and \HeII\ 256~\AA\ (as
labelled). The \HeII\ images are scaled from left to right (b) linearly (c)
logarithmically (d) relative to the average for the one hour observation
period. The three explosive events, indicated with arrows and seen
on the right of the \HeII\ frames, are
blue shifts because the spectral images have been flipped so that west is on
the right. All images are shown as negatives.}
    \label{heiishifts}
   \end{figure}

Figure~\ref{heiishifts} demonstrates the method used to enhance blue shifts in
\HeII\ at the edge of the image. Here images from a region of quiet Sun in \FeXII\
(a) and \HeII\ (b-d) are shown. There is no structure in the \FeXII\ but the
\HeII\ (b) shows the outline of supergranular cells. To bring out the
brightening on the west edge (blue wing) the logarithm of intensity is shown
in (c) and finally in (d) the image divided by the average \HeII\ image for
the one hour period of observation. At this time there were three explosive
events around sun-x = -70\arcsec, -45\arcsec, and -20\arcsec. It is not
possible to determine the Doppler velocity precisely because these are not real
line profiles. We estimate that they are of the order of a couple of pixels
or 60~\kms.
Another advantage
of dividing by the average for the observing period is that brightenings
in the image would also show up and warn that edge brightenings may be
due to a neighbouring line brightening in a structure inside the image. This is not
the case for the three events identified.

\begin{figure}
\centering
\includegraphics[width=11.0 cm]{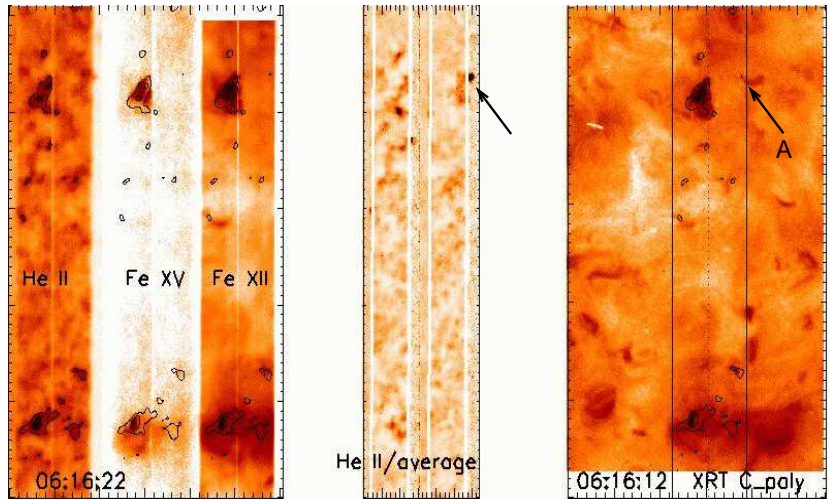}
\caption{A frame from the movie showing the relationship
 between \HeII, \FeXV, \FeXII, and \HeII\ lineshifts and
X-ray emission from a region of quiet Sun on 7 November 2007.
The left panel shows \HeII, \FeXV\ and \FeXII\ EIS wide slit images
from two raster positions patched together, as linearly scaled negatives
with contours of bright \HeII\ emission overlaid. The center panel shows
the two neighbouring \HeII\ images divided by their average \HeII\ image for the period.
Here, the windows are not patched together so that emission changes beyond the
image edge along the center line can be seen.
Doppler shift
events show up as black regions beyond the image edges
which are marked by solid white lines.
 An arrow points to
a Doppler shift event near X-ray loop A. The right panel shows the
closest in time XRT image with \HeII\ contours overlaid.}
    \label{fmovsnap}
   \end{figure}

The advantage of the wide slit is that it shows the structure surrounding
these Doppler shift events. We have therefore combined the \HeII, \FeXII\ and
\FeXV\ wide slit images with the Doppler shift information in a short movie,
showing one hour of quiet Sun observations. One frame from the movie is shown
in Figure~\ref{fmovsnap}. During the observing sequence two wide slit positions,
separated by 40\arcsec\ were alternated, giving an 80\arcsec\ image every
1.5~min. The closest XRT C-poly image is placed alongside with \HeII\
contours overlaid. It is striking how often bright \HeII\ is found near the
edge (footpoints) of X-ray structures.  One good example is the structure, marked A, on
Figure~\ref{fmovsnap}, where some of the strongest \HeII\ shifts were seen.
In general the \FeXII\ and X-ray emissions
correlate well.

\begin{figure}
\centering
\includegraphics[width=14.0 cm]{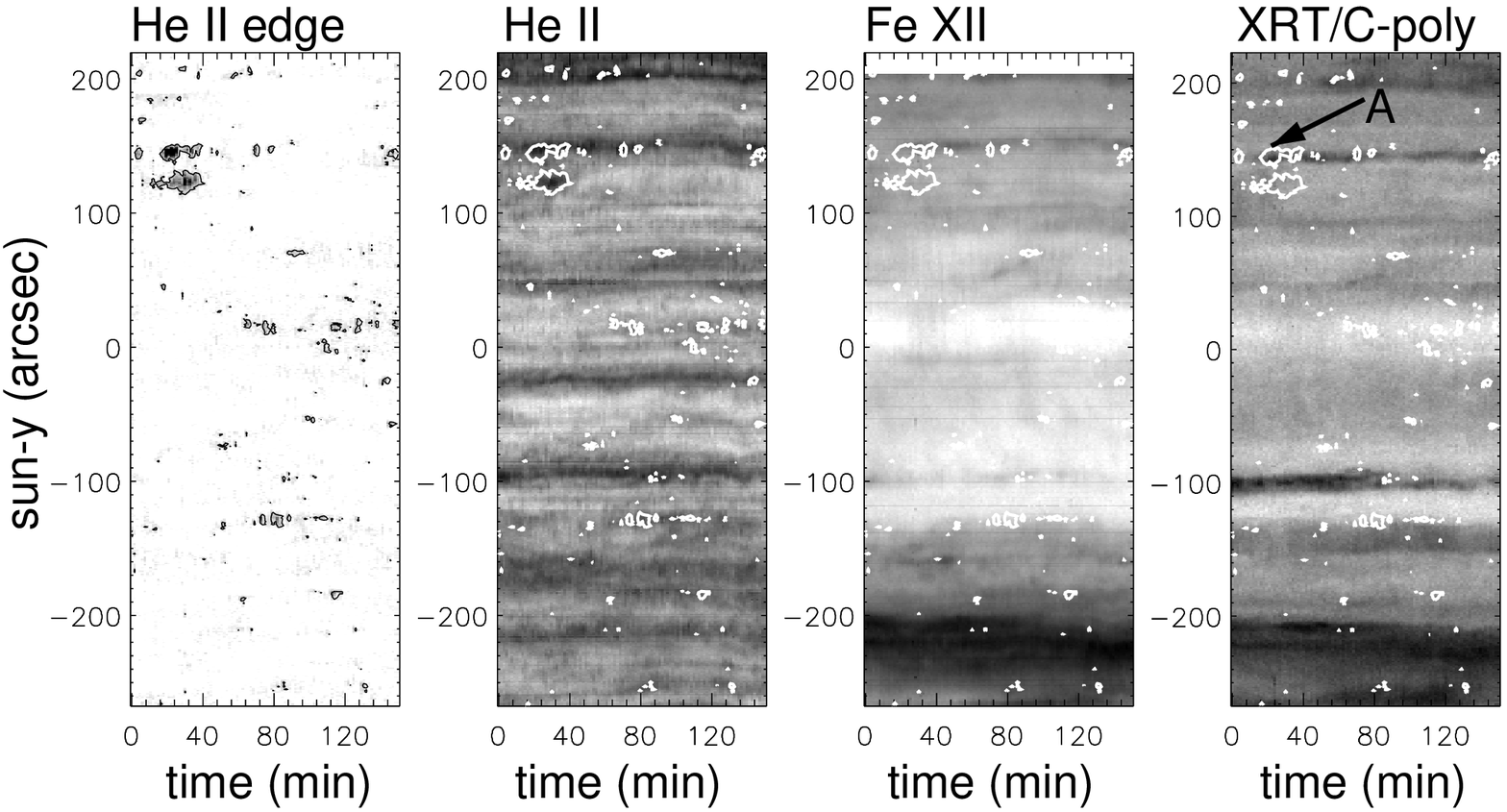}
\caption{Time series of (a) \HeII\ intensity beyond the west edge of the
image (blue shifts), (b) \HeII\ and (c) \FeXII\ intensity at the west edge
and (d) time series of X-ray intensity at the position of the west edge. The
contours outline positions of \HeII\ intensity beyond the image edge. In (a)
darkness indicates the increase in blue wing intensity
above the average for that
position along the slit and (b) and (c) are linearly scaled intensity negatives
 and (d)
is a logarithmically scaled intensity negative. The time is minutes after 06:05:16 UT
on 7 November 2007. }
    \label{timeshifts}
   \end{figure}

 Time series of emission beyond the image edge of \HeII\ and just
inside  \HeII\ and \FeXII\ images are shown in
Figure~\ref{timeshifts} alongside the X-ray emission from the position of the
images' edge. Any \HeII\ emission seen off the main image is produced by
Doppler shifts from a region inside the image, so this directly compares
\HeII\ shifts with the \HeII\ intensity and hotter plasma (\FeXII\ and X-ray)
emission. The \HeII\ is reminiscent of explosive event distributions, where
there are preferred event sites along network lanes spaced
$50-100$\arcsec\ apart, with events lasting 2-3~min, sometimes seen in
bursts of up to 30~min \citep{IBGW97a, Innes01}. \HeII\ shifts seem to occur
more frequently near sites of enhanced \FeXII\ and X-ray emission.

\subsection{High Doppler Shifts near X-ray Spot Brightening}
\begin{figure}
\centering
\includegraphics[width=9 cm]{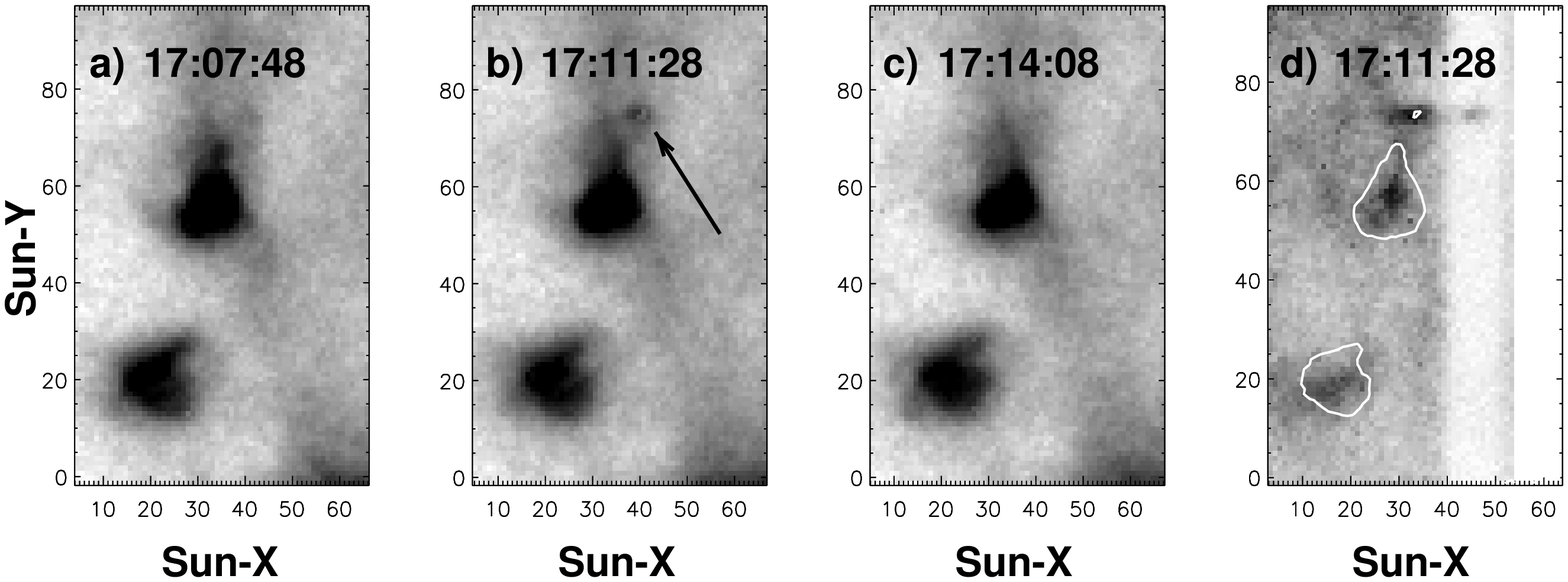}
\caption{Time series of XRT Al-mesh intensity showing hot spot brightening
(indicated with an arrow) and (d) the EIS \SiVII\ image with XRT contours
on 10 April 2007.  All intensities are
represented as logarithmically scaled negatives. }
    \label{xspot}
   \end{figure}

\begin{figure}
\centering
\includegraphics[width=9 cm]{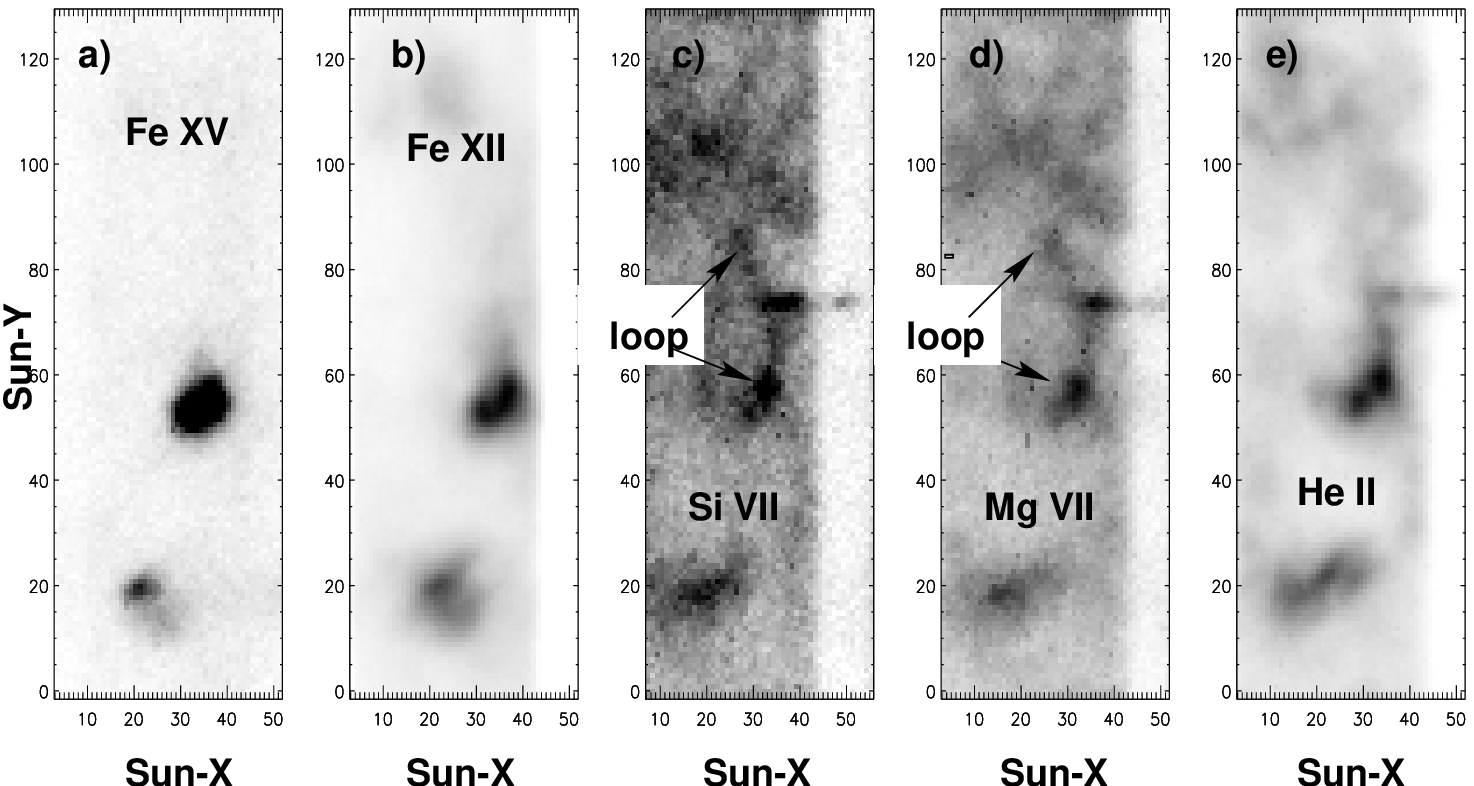}
\caption{Upper transition region plasmoid or jet seen at 17:11:15 UT on 10 April 2007.
 The
strong blue shift event is seen on the right hand side of \SiVII\ 275.35~\AA,
\MgVII\ 278.44~\AA, and \HeII. All spectral images
show blue shifts on the right, and are
represented as linearly scaled negatives. }
    \label{trshifts}
   \end{figure}

 This observation, illustrated in Figs.~\ref{xspot} and \ref{trshifts},
is very unusual because the edge of the wide slit caught the brightening of
an intense X-ray spot. Figure~\ref{xspot}a-c, shows the time sequence of X-ray
images. The X-ray brightening lasts less than 7~min. Simultaneously, there is
a horizontal streak in all transition region and lower coronal line images
with a length corresponding to a line width of 220~\kms\ beyond the window
edge (Figure~\ref{trshifts}). There is no signature in the EIS coronal lines
\FeXV\ and \FeXII.

The hot spot seems to be coming from the top of a cool loop seen faintly in
the \SiVII\ and \MgVII\ images. The images in the hotter lines, \FeXV\ and
\FeXII\ and the blends in \HeII, suggest that one end of the cool coronal
loop is rooted in a bright point and the other in the quiet Sun. The event
lasted 1.5~min in \SiVII\ and \MgVII\ and continued for another 3.5~min in
the cooler \HeII\ line. At first glance the structure has the classic
configuration of a reconnection jet above a coronal loop \citep[cf][]{Hira74,
KP76}.

\subsection{\FeXV\ Doppler Shift}

\begin{figure}
\centering
\includegraphics[width=11.0 cm]{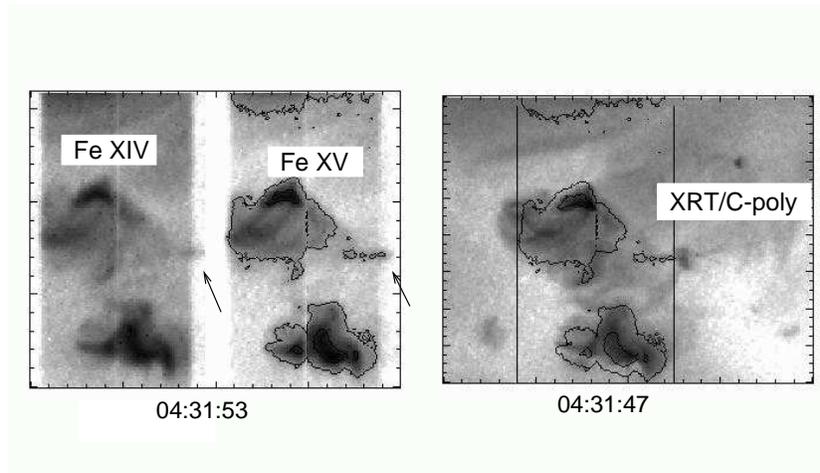}
\caption{Example of Doppler shift seen in coronal lines on 15 November 2007.
 The images, from left to right,
are EIS \FeXIV, \FeXV\ and XRT C-poly. Arrows point to Doppler shifts seen beyond
the image edge. The EIS image edges are indicated with solid
lines in the XRT image. The contours outline the enhanced \FeXV\ emission.}
    \label{nov15event}
   \end{figure}

Shifts in coronal lines with formation temperature about $2\times10^6$~K are
also occasionally seen from X-ray structures. One example of a
 hot \FeXV\ and \FeXIV\ blue shift of about 120~\kms\
  is shown in Figure~\ref{nov15event}.
 The outflow comes from an almost horizontal X-ray structure
 connecting a small X-ray
 brightening to a larger, more persistent X-ray region. It could be flow
 along a hot loop connecting the two X-ray regions.
 The shifts last
  about 25~min, during which time the brightening was seen to flare up and
 fade.
The shift is also seen in \FeXII\ over a shorter period of time and with
lower velocity.

\section{Discussion}
There is a wealth of information in the EIS 40\arcsec\ wide slit images. The primary
use of wide slit images is to see rapid time variations of structures
in specific spectral lines \citep{Hansteen07}, typically with a cadence 20 times
(the slit width) faster than rastering with the 2~\arcsec\ slit. After passing through
the slit, the light is dispersed producing rows of overlapping spectral images.
 If a Doppler shift
occurs in the center of the image, this is indistinguishable from a horizontal structure.
It is therefore beneficial to have simultaneous filtergrams to show the
underlying structures.

Here we show how the wide slit can be used
to detect Doppler shifts at the edge of the image.
We have highlighted three types of Doppler shift event seen in EIS wide slit
spectra: explosive events in \HeII\ at the image edge; a fast jet related to
an X-ray spot brightening; and \FeXV\ flow connecting  an X-ray spot
brightening to a larger X-ray region. The latter observations were done in
sit-and-stare at two positions spaced 40\arcsec\ apart in order to pick up first the
blue and then the red wings of the lines.
At present our understanding of the line profile at the edge of the slit is not
sufficiently well understood to determine accurate line shifts but flows producing
emission more than two pixels away from the average line profile at the
edge are clearly detectable.

  This suggests new
ways of investigating, for example, explosive event dynamics because one can
see the loop motion and heating beyond the position of the explosive event.
It could also be
 useful when investigating Doppler shift fluctuations
from hot loops because narrow slit observations in sit-and-stare are
 not able to distinguish real intensity fluctuations at a point from
movement of loops into and out of the slit field-of-view
\citep[cf][]{Wetal03b, Mariska07}.

 The aim of this short paper is to show the
possibilities for future studies, not to present event details and their
scientific implications. Discussions on the events themselves, including
detailed analyses from SOT on \hinode\, and SUMER on SOHO will be given
in more extensive papers.

\begin{acknowledgements}
We would like to thank Khalid Al-Janabi for his help with the EIS
studies. Also many thanks to the referee for their constructive comments.
\hinode\ is a Japanese mission developed and launched by ISAS/JAXA,
collaborating with NAOJ, as domestic partner, and NASA (USA) and STFC (UK)
as international partners. Scientific operation of the \hinode\ mission is
conducted by the \hinode\ science team organized at ISAS/JAXA.
Support for the postlaunch operation is provided by JAXA and NAOJ, STFC,
NASA, ESA (European Space Agency), and NSC (Norway).
We are grateful to all teams for their efforts in the design, building, and
operation of the \hinode\ and SoHO missions.
\end{acknowledgements}

\bibliographystyle{aa}

\begin{thebibliography}{20}
\expandafter\ifx\csname natexlab\endcsname\relax\def\natexlab#1{#1}\fi

\bibitem[{{Brueckner} \& {Bartoe}(1983)}]{BB83}
{Brueckner}, G.~E. \& {Bartoe}, J.-D.~F. 1983, Astrophys. J., 272, 329

\bibitem[{{Chae} {et~al.}(1998){Chae}, {Wang}, {Lee}, {Goode}, \&
  {Sch{\"u}hle}}]{Cetal98}
{Chae}, J., {Wang}, H., {Lee}, C., {Goode}, P.~R., \& {Sch{\"u}hle}, U. 1998,
  Astrophys. J., 504, L123

\bibitem[{{Culhane} {et~al.}(2007){Culhane}, {Harra}, {James}, {Al-Janabi},
  {Bradley}, {Chaudry}, {Rees}, {Tandy}, {Thomas}, {Whillock}, {Winter},
  {Doschek}, {Korendyke}, {Brown}, {Myers}, {Mariska}, {Seely}, {Lang}, {Kent},
  {Shaughnessy}, {Young}, {Simnett}, {Castelli}, {Mahmoud}, {Mapson-Menard},
  {Probyn}, {Thomas}, {Davila}, {Dere}, {Windt}, {Shea}, {Hagood}, {Moye},
  {Hara}, {Watanabe}, {Matsuzaki}, {Kosugi}, {Hansteen}, \&
  {Wikstol}}]{Culhane07}
{Culhane}, J.~L., {Harra}, L.~K., {James}, A.~M., {et~al.} 2007, Solar Phys.,
  60

\bibitem[{{Dere} {et~al.}(1989){Dere}, {Bartoe}, \& {Brueckner}}]{DBB89}
{Dere}, K.~P., {Bartoe}, J.-D.~F., \& {Brueckner}, G.~E. 1989, Solar Phys.,
  123, 41

\bibitem[{{Golub} {et~al.}(2007){Golub}, {Deluca}, {Austin}, {Bookbinder},
  {Caldwell}, {Cheimets}, {Cirtain}, {Cosmo}, {Reid}, {Sette}, {Weber},
  {Sakao}, {Kano}, {Shibasaki}, {Hara}, {Tsuneta}, {Kumagai}, {Tamura},
  {Shimojo}, {McCracken}, {Carpenter}, {Haight}, {Siler}, {Wright}, {Tucker},
  {Rutledge}, {Barbera}, {Peres}, \& {Varisco}}]{Golub07}
{Golub}, L., {Deluca}, E., {Austin}, G., {et~al.} 2007, Solar Phys., 243, 63

\bibitem[{{Hansteen} {et~al.}(2007){Hansteen}, {de Pontieu}, {Carlsson},
  {McIntosh}, {Watanabe}, {Warren}, {Harra}, {Hara}, {Tarbell}, {Shine},
  {Title}, {Schrijver}, {Tsuneta}, {Katsukawa}, {Ichimoto}, {Suematsu}, \&
  {Shimizu}}]{Hansteen07}
{Hansteen}, V.~H., {de Pontieu}, B., {Carlsson}, M., {et~al.} 2007, PASJ, 59,
  699

\bibitem[{{Hirayama}(1974)}]{Hira74}
{Hirayama}, T. 1974, Solar Phys., 34, 323

\bibitem[{{Innes}(2001)}]{Innes01}
{Innes}, D.~E. 2001, Astron. Astrophys., 378, 1067

\bibitem[{{Innes} {et~al.}(1997a){Innes}, {Brekke}, {Germerott}, \&
  {Wilhelm}}]{IBGW97a}
{Innes}, D.~E., {Brekke}, P., {Germerott}, D., \& {Wilhelm}, K. 1997a, Solar
  Phys., 175, 341

\bibitem[{{Innes} {et~al.}(1997b){Innes}, {Inhester}, {Axford}, \&
  {Wilhelm}}]{IIAW97b}
{Innes}, D.~E., {Inhester}, B., {Axford}, W.~I., \& {Wilhelm}, K. 1997b, Nat,
  386, 811

\bibitem[{{Kopp} \& {Pneuman}(1976)}]{KP76}
{Kopp}, R.~A. \& {Pneuman}, G.~W. 1976, Solar Phys., 50, 85

\bibitem[{{Mariska} {et~al.}(2007){Mariska}, {Warren}, {Ugarte-Urra}, {Brooks},
  {Williams}, \& {Hara}}]{Mariska07}
{Mariska}, J.~T., {Warren}, H.~P., {Ugarte-Urra}, I., {et~al.} 2007, PASJ, 59,
  713

\bibitem[{{Ning} {et~al.}(2004){Ning}, {Innes}, \& {Solanki}}]{NIS04}
{Ning}, Z., {Innes}, D.~E., \& {Solanki}, S.~K. 2004, Astron. Astrophys., 419,
  1141

\bibitem[{{Tsuneta} {et~al.}(2008){Tsuneta}, {Ichimoto}, {Katsukawa}, {Nagata},
  {Otsubo}, {Shimizu}, {Suematsu}, {Nakagiri}, {Noguchi}, {Tarbell}, {Title},
  {Shine}, {Rosenberg}, {Hoffmann}, {Jurcevich}, {Kushner}, {Levay}, {Lites},
  {Elmore}, {Matsushita}, {Kawaguchi}, {Saito}, {Mikami}, {Hill}, \&
  {Owens}}]{Tsuneta08}
{Tsuneta}, S., {Ichimoto}, K., {Katsukawa}, Y., {et~al.} 2008, Solar Phys.,
  249, 167

\bibitem[{{Wang} {et~al.}(2003){Wang}, {Solanki}, {Innes}, {Curdt}, \&
  {Marsch}}]{Wetal03b}
{Wang}, T.~J., {Solanki}, S.~K., {Innes}, D.~E., {Curdt}, W., \& {Marsch}, E.
  2003, Astron. Astrophys., 402, L17

\bibitem[{{Wilhelm} {et~al.}(1995){Wilhelm}, {Curdt}, {Marsch}, {Sch{\"u}hle},
  {Lemaire}, {Gabriel}, {Vial}, {Grewing}, {Huber}, {Jordan}, {Poland},
  {Thomas}, {Kuhne}, {Timothy}, {Hassler}, \& {Siegmund}}]{Wetal95}
{Wilhelm}, K., {Curdt}, W., {Marsch}, E., {et~al.} 1995, Solar Phys., 162, 189

\bibitem[{{Wilhelm} {et~al.}(1998){Wilhelm}, {Innes}, {Curdt}, {Kliem}, \&
  {Brekke}}]{Wilhelm98}
{Wilhelm}, K., {Innes}, E.~E., {Curdt}, W., {Kliem}, B., \& {Brekke}, P. 1998,
  in ESA Special Publication, Vol. 421, Solar Jets and Coronal Plumes, ed.
  T.-D. {Guyenne}, 103

\bibitem[{{Winebarger} {et~al.}(2002){Winebarger}, {Emslie}, {Mariska}, \&
  {Warren}}]{WEMW02}
{Winebarger}, A.~R., {Emslie}, A.~G., {Mariska}, J.~T., \& {Warren}, H.~P.
  2002, Astrophys. J., 565, 1298

\bibitem[{{Young} {et~al.}(2007a){Young}, {Del Zanna}, {Mason}, {Dere},
  {Landi}, {Landini}, {Doschek}, {Brown}, {Culhane}, {Harra}, {Watanabe}, \&
  {Hara}}]{Young07a}
{Young}, P.~R., {Del Zanna}, G., {Mason}, H.~E., {et~al.} 2007a, PASJ, 59, 857

\bibitem[{{Young} {et~al.}(2007b){Young}, {Del Zanna}, {Mason}, {Doschek},
  {Culhane}, \& {Hara}}]{Young07b}
{Young}, P.~R., {Del Zanna}, G., {Mason}, H.~E., {et~al.} 2007b, PASJ, 59, 751

\end{thebibliography}

\end{document}